\documentclass[a4paper,fleqn,usenatbib]{mnras}

\usepackage[T1]{fontenc}
\usepackage{ae,aecompl}

\newcommand{\lsim}{\mathrel{\rlap{\raise -.3ex\hbox{${\scriptstyle\sim}$}}%
		  \raise .6ex\hbox{${\scriptstyle <}$}}}%
\newcommand{\gsim}{\mathrel{\rlap{\raise -.3ex\hbox{${\scriptstyle\sim}$}}%
		  \raise .6ex\hbox{${\scriptstyle >}$}}}%

\usepackage{graphicx}	
\usepackage{amsmath}	
\usepackage{amssymb}	

\title[Space objects are a new skyglow threat]{The proliferation of space objects is a rapidly increasing source of artificial night sky brightness}

\author[M. Kocifaj et al.]{
M.~Kocifaj,$^{1,2}$\thanks{E-mail: kocifaj@savba.sk}
F.~Kundracik,$^{2}$
J.~C.~Barentine,$^{3,4}$
S.~Bar\'{a},$^{5}$
\\
$^{1}$ICA, Slovak Academy of Sciences, D\'{u}bravsk\'{a} cesta 9, 845 03 Bratislava, Slovakia\\
$^{2}$Department of Experimental Physics, FMPI, Comenius University, Mlynsk\'{a} dolina, 842 48 Bratislava, Slovakia\\
$^{3}$International Dark-Sky Association, 3223 N. 1st Ave, Tucson, AZ 85710 USA\\
$^{4}$Consortium for Dark Sky Studies, University of Utah, 375 S 1530 E, RM 235 ARCH,
Salt Lake City, Utah 84112-0730 USA\\
$^{5}$Departamento de F\'{\i}sica Aplicada, Universidade de Santiago de Compostela, 
E-15782 Santiago de Compostela, Galicia, Spain
}

\date{Accepted XXX. Received YYY; in original form ZZZ}

\pubyear{2015}

\begin{document}
\label{firstpage}
\pagerange{\pageref{firstpage}--\pageref{lastpage}}
\maketitle

\begin{abstract}
The population of artificial satellites and space debris orbiting the Earth imposes non-negligible constraints on both space operations and ground-based optical and radio astronomy. The ongoing deployment of several satellite `mega-constellations' in the 2020s represents an additional threat that raises significant concerns. The expected severity of its unwanted consequences is still under study, including radio interference and information loss by satellite streaks appearing in science images. In this Letter, we report a new skyglow effect produced by  space objects: increased night sky brightness caused by sunlight reflected and scattered by that large set of orbiting bodies whose direct radiance is a diffuse component when observed with the naked eye or with low angular resolution photometric instruments. According to our preliminary estimates, the zenith luminance of this additional light pollution source may have already reached $\sim$20 $\mu$cd m$^{-2}$, which amounts to an approximately 10 percent increase over the brightness of the night sky determined by natural sources of light. This is the critical limit adopted in 1979 by the International Astronomical Union for the light pollution level not to be exceeded at the sites of astronomical observatories.
\end{abstract}



\section{Introduction}
Artificial satellites orbiting the Earth have been a concern of astronomers since the 
launch of the first such object, Sputnik 1, in 1957. As of 1 January 2021, some 
3,372 satellites are in orbit~\citep[][]{UCS2021} along with many tens of thousands of 
pieces of space debris; we refer to both satellites and space debris here as ``space 
objects.''

The orbital altitudes of space objects range from a few hundred kilometres in 
the case of objects in Low-Earth Orbit (LEO) to beyond the 35,786-km height defining 
geosynchronous orbits. At such altitudes, space objects remain directly illuminated by 
sunlight as seen from the night side of the Earth; consequently, they appear in images 
obtained with ground-based telescopes as streaks of various lengths and apparent 
brightness depending on the orbital parameters of the objects. Because the streaks are 
often comparable to or brighter than objects of astrophysical interest, their presence 
tends to compromise astronomical data and poses the threat of irretrievable loss of 
information.

The number of space objects orbiting Earth is expected to increase by more than an 
order of magnitude in the next decade due to the launches of fleets of new, large 
'constellations' of communications satellites. While astronomers can often plan 
observations around the presence of bright space objects in their telescopes' fields of
view by predicting their apparent positions on the night sky using databases of orbital
elements, the expected increase in the space object population significantly increases 
the probability that any particular observation will be affected by streaks. 
\citet[][]{HainautEtAl2020} and \citet[][]{Tyson2020} recently explored the expected
impacts of large satellite constellations on both optical and near-infrared 
astronomical observations, finding that ultra-wide imaging exposures made with large 
survey telescopes will be most affected, with up to 40\% of science exposures affected 
by streaks assuming the largest expected numbers of new satellites and an overall loss 
of $\sim$1\% of science pixels.

To date there is little information in the literature as to the contribution of space 
objects to the diffuse brightness of the night sky, as opposed to the effect of discrete streaks that add systematic errors to astronomical data. While the night sky luminance 
signal due to sunlight reflected from space objects and propagating in the Earth's 
atmosphere is expected to be small, it is otherwise unknown how its contribution to 
diffuse night sky brightness (NSB) compares to natural sources of light in the night sky 
that ultimately determine the sky luminance over `pristine' locations nearly devoid of 
anthropogenic skyglow, such as the sites of astronomical research observatories.

We decided to estimate the present-day diffuse NSB contribution from space objects, before the ongoing
deployment of large communication satellite constellations increase in order to benchmark
future NSB observations in these pristine places. In the current work, we describe the 
geometry of the situation and model the diffuse NSB specifically attributable to sunlight
reflected from space objects, establishing lower limits for the space object contribution
based on expectations for coming launches and the potential for orbital crowding to 
generate new sources of space debris.

This paper is organised as follows. We develop the requisite theory, establish the 
expected spectral radiance of space objects in the observer's zenith, and compute the 
zenith night sky luminance contribution from space objects in Section~\ref{sec:theory}. 
We present the results of our calculations and discuss their implications in 
Section~\ref{sec:results}. Finally, we draw conclusions from this work and speculate on 
future prospects in Section~\ref{sec:conclusions}.

\begin{figure}
	\includegraphics[width=\columnwidth]{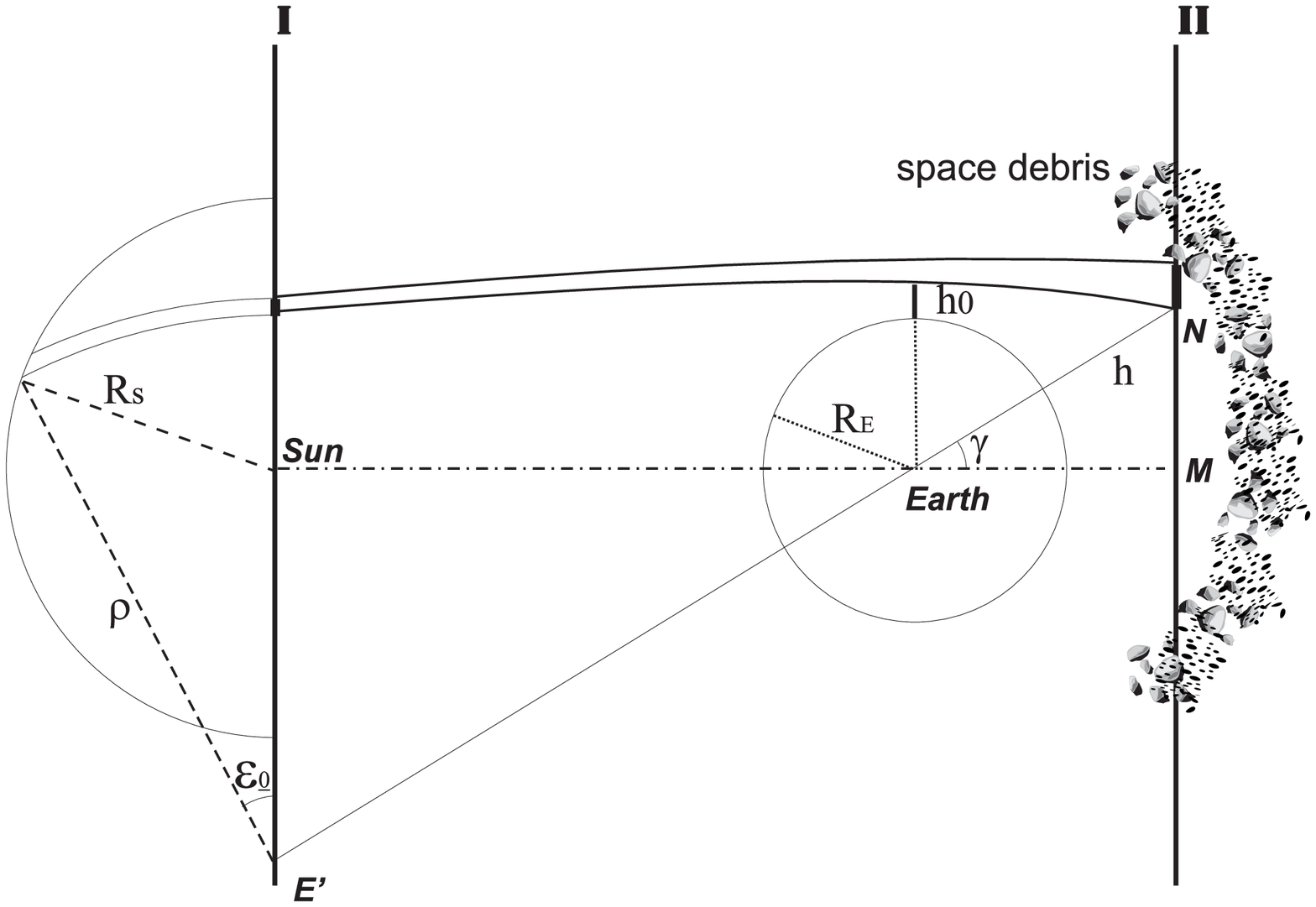}
    \caption{A simplified illustrative view of space debris orbiting 
    the Earth and directly illuminated by sunlight. Planes I and II are for the Sun and 
    the Earth, respectively, while the parameters used are described in the main text.
    The drawing is not to any particular scale.}
   \label{fig:geometry}
\end{figure}

\section{Sky brightness due to space objects}
\label{sec:theory}
Due to their high altitudes, objects in orbit around the Earth can be visible after 
the onset of astronomical night; i.e., for solar depression angles $\geq$18$^{\circ}$. For
instance, objects at altitudes between 1000--2000 kilometres can be directly illuminated 
by sunlight and observed in the zenith even when the depression angle of the solar disc is
30-40$^{\circ}$. The lower the minimum altitude of light rays, the more such rays diverge 
due to atmospheric refraction; see the quantity $h_{0}$ in Fig.~\ref{fig:geometry}. 
Depending on beam trajectory, the angle of refraction is 1\arcmin~at an altitude of 30 km,
10\arcmin ~at $h_{0} \approx$ 15 km, and approaches 1$^{\circ}$ for rays entering the 
lower troposphere~\citep[][]{Link1969}. Satellites illuminated by such low rays are 
normally dark or even invisible because of exceptional atmospheric extinction of sunlight.

Since the deflection of the light beam is extremely low, most space objects orbiting the 
Earth are either directly illuminated by sunlight or disappear once they cross into the 
geometrical shadow of the Earth.\footnote{The geometrical shadow of the Earth neglects 
refraction effects from the terrestrial atmosphere.} A few objects moving in a `transition
zone' between positions unaffected by the Earth's atmosphere and those obstructed by the 
Earth contribute negligibly to the diffuse brightness of the night sky, and thus are not 
considered in the following analysis. The transition zone is defined by the 
angle of refraction, which normally does not exceed 1$^{\circ}$. The objects traversing
the sky from the twilight region into the Earth's shadow are constantly illuminated by sunlight and 
seen in the visual field of 90$^{\circ}$ or more; thus, the number of objects in the 
transition zone is about 1\% of all objects we can see in the night sky.

\subsection{Spectral radiance}
\label{sec:radiance}
The spectral irradiance (W m$^{-2}$ nm$^{-1}$) at a point $N$ located at an altitude 
$h$ above the Earth's surface is 
\begin{eqnarray}\label{eq:irradiance}
    E_{\lambda}(\gamma) &=& 2 b_{0,\lambda} \int_{\gamma-R_{S}}^{\gamma+R_{S}} 
	T_{\lambda}^{Ext}(\rho)~\epsilon_{0}(\rho)~d\rho~,
\end{eqnarray}
where $b_{0,\lambda}$ is the average radiance of the Sun (W m$^{-2}$ 
nm$^{-1}$ sr$^{-1}$), $\lambda$ is the wavelength of radiation, $\gamma$ is the angular 
distance measured from the centre of the Earth between the position of 
an illuminated object (in point $N$) and the centre of the Earth's shadow (in point $M$), 
$R_{S}$ is the angular radius of the solar disc, $\rho$ is the angular distance ranging 
from $\gamma-R_{S}$ to $\gamma+R_{S}$,and $\epsilon_{0}$ is the maximum opening angle at 
an angular distance of $\rho$, i.e.,the angle between centre of solar disc and the limb 
(consult Fig.~\ref{fig:geometry}). The transmission coefficient $T_{\lambda}^{Ext}$ 
characterizes attenuation of sunlight in the atmosphere due to extinction~\citep[see, 
e.g.,][]{KocifajHorvath2005}. In the absence of the terrestrial atmosphere, or for 
high-altitude beams with $h_{0} \gsim$ 60 km~\citep[][]{Link1969}, 
Eq.~(\ref{eq:irradiance}) reduces to
\begin{eqnarray}\label{eq:irradiance0}
    E_{0,\lambda} &=& 2 b_{0,\lambda} \int_{\gamma-R_{S}}^{\gamma+R_{S}} 
	~\epsilon_{0}(\rho)~d\rho~= b_{0,\lambda} \pi R_{S}^{2}~,
\end{eqnarray}
with $\pi R_{S}^{2}$ being the solid angle subtended by the solar disc in steradians. 

Of all photons entering the point $N$, a small fraction is intercepted by space objects and redirected toward the Earth. The redirection is dominated by scattering in the case of debris objects smaller than or comparable to the wavelengths of visible light. For larger macroscopic objects to which the principles of geometric optics are applicable, direct reflection of light dominates the redirection. The amount of redirection is proportional to $E_{0,\lambda} \frac{P_{\lambda}(\theta)}{4\pi} k_{\lambda}(h)$, where $P_{\lambda}(\theta)$ is scattering phase function normalized to $4\pi$, and $\theta$ is the scattering angle, i.e., the angle between the incident beam and the scattered beam. Conservation of energy requires that $\int \frac{P_{\lambda}(\theta)}{4\pi} d\omega=1$, where $d\omega=2\pi \sin(\theta) d\theta$ is the elementary solid angle. 

Artificial satellites and pieces of space debris have diverse shapes and
are oriented randomly in the line of sight, scattering and reflecting light in a complex 
manner. However, the number of orbital objects is large, and statistical averaging over a
large sample size yields results similar to those for equally sized spherical objects. 
Since most space objects are very large compared to the wavelengths of visible light, 
they scatter light strictly in the geometrical optics regime. The resulting optical 
signal therefore depends much more strongly on an object's geometric cross section rather
than on its morphology. 

The volume scattering coefficient $k_{\lambda}(h)$ can be computed from Mie 
theory~\citep[see, e.g.,][]{HergertWriedt2012} by treating a given object as a sphere of 
radius of $r$:
\begin{eqnarray}\label{eq:Mie}
    k_{\lambda}(h) &=& \pi \int_{0}^{\infty} r^{2} n(r,h)~Q_{\lambda,sca}(r)~dr,
\end{eqnarray}
where $Q_{\lambda,sca}(r)$ is the scattering efficiency factor for an object of radius $r$
illuminated by monochromatic light of wavelength $\lambda$. The function $n(r,h) dr$ is 
the number of objects of radius $r \rightarrow r+dr$ per cubic meter such that $n(r,h)$ is
expressed in units of $m^{-4}$. For objects larger than the typical observational 
wavelengths $Q_{\lambda,sca} \approx$ 1 to 2. 

\begin{figure}
	\includegraphics[width=\columnwidth]{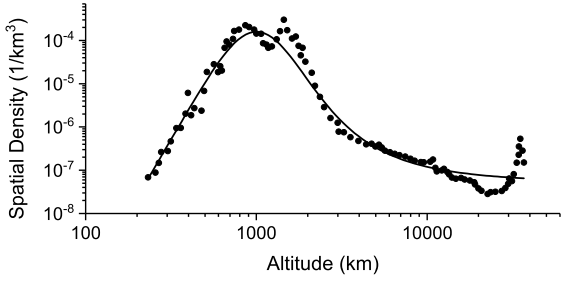}
    \caption{Spatial density of space objects larger than 1 mm as a function 
	of altitude. Dots are data published by~\citet[][]{Bendisch2004}, while the solid 
	line represents our fitted model.}
   \label{fig:stratification}
\end{figure}

The spectral radiance due to the reflection of sunlight in all space objects in Earth orbit along the line of sight $z$, detected by an observer at ground level is
\begin{eqnarray}\label{eq:radiance}
    L_{\lambda}(z) &=& \frac{e^{-\tau_{\lambda}/\cos z}}{\cos z} E_{0,\lambda}
	\int_{h_{1}}^{\infty} \frac{P_{\lambda}(\theta)}{4\pi} k_{\lambda}(h)~dh,
\end{eqnarray}
where $\tau_{\lambda}$ is the optical thickness of Earth's atmosphere, $z$ is the zenith 
angle of observation, and $h_{1}$ is the minimum orbital altitude of space objects. The 
latter parameter is normally a few hundred kilometers (see Fig.~\ref{fig:stratification}),
but the lower integration limit can also be asymptotically extended to $h_{1}=0$ km 
assuming that both the number density of space objects and $k_{\lambda}(h)$ approach zero
near the Earth's surface. Space objects with arbitrary shapes reflect sunlight more or 
less isotropically, resulting in $P_{\lambda}(\theta)/4\pi \approx 
1/4\pi$~\citep[][]{Shendeleva2017}. Substituting the approximate expressions into 
Eq.~(\ref{eq:radiance}), we found for the theoretical spectral radiance in the zenith 
($z=0^{\circ}$)
\begin{eqnarray}\label{eq:radiance_approx}
    L_{\lambda}(z) &=& \frac{E_{0,\lambda} e^{-\tau_{\lambda}}}{2}
	\int_{h=h_{1}}^{\infty} \int_{r=0}^{\infty} r^{2}~n(r,h)~dr~dh.
\end{eqnarray}

\subsection{Zenith sky luminance: An estimate}
\label{sec:luminance}
\begin{figure}
	\includegraphics[width=\columnwidth]{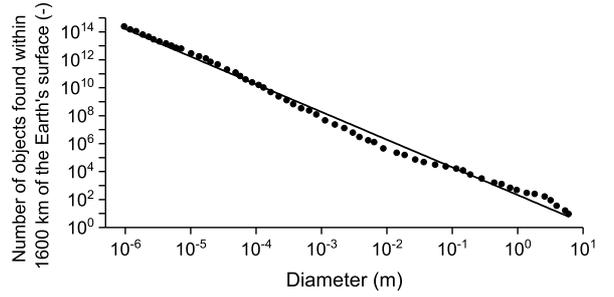}
    \caption{Logarithmic relationship between cumulative number and sizes of LEO objects.
	Dots are data published by \citet[][]{NRC1995} while the solid line is the analytic 
	fit. The data shown are for objects within 1600 kilometres of the Earth's surface.}
   \label{fig:cumulative}
\end{figure}

It follows from Eq.~(\ref{eq:radiance_approx}) that the number distribution of space 
objects is the only otherwise unknown quantity required to calculate these objects' 
contribution to total diffuse night sky brightness. We combined a number of works
to estimate this function, assuming $n(r,h)=R(r) H(h)$. The function $R(r)$ models
the size distribution (m$^{-1}$), while $H(h)$ is the number concentration of space
objects (number m$^{-3}$). The cumulative number of LEO objects 
larger than a given size has been known for decades~\citep[][]{NRC1995}. The rate at 
which the cumulative number of objects, $N(r)$, declines with increasing size is roughly 
constant on a log scale, which allows for the simple logarithmic approximation $\log N(r) 
= a \log(2r)+b$ (Fig.~\ref{fig:cumulative}), where $2r$ is the characteristic diameter of 
objects in meters, and $a=-1.98$ and $b=2.32$ are dimensionless scaling constants. 

In Fig.~\ref{fig:cumulative}, $N(r)=n_{0} \int_{r}^{5 m} R(r)dr$, where $n_{0} = 1.64 \times 10^{14}$ is the dimensionless coefficient of proportionality satisfying the normalization condition 
$\int_{5 \times 10^{-7}m}^{5m} R(r) dr=1$.
The function $H(h)$ is scaled to satisfy the following equation
\begin{eqnarray}\label{eq:n0}
	\int_{200km}^{1600km} 4 \pi h^2 H(h)~dh &=& n_{0} .
\end{eqnarray}
The integral form for $N(r)$ can be easily transformed into 
differential form $R(r)=-\frac{1}{n_{0}} \frac{dN(r)}{dr}=-\frac{a}{r~n_{0}} 10^{a\log(2r)+b}$.
The vertical stratification of objects in Earth orbit was modeled in accordance with 
\citet[][]{Bendisch2004}. We have developed an analytical formula that models the 
experimental data reasonably well for altitudes $h$ < 40000 km 
(Fig.~\ref{fig:stratification}). Assuming $h$ is given in metres, we found 
\begin{eqnarray}\label{eq:Hh}
    H(h) &=& \beta ~10^{A(\log h - D)/[B+(\log h - F)^{m}]-C}~
\end{eqnarray}
with $A$=4.02, $B$=0.60, $C$=7.3, $D$=5.35, $F$=5.3, and $m$=5.32. The coefficient of 
proportionality, $\beta$=0.10, was derived from the normalization condition  
(Eq.~\ref{eq:n0}).

To estimate the contribution of space objects to the diffuse brightness of the night sky 
at the zenith we take $E_{0,\lambda} e^{-\tau_{\lambda}}$ to be about $E_{0,vis} 
\approx$ 120,000 lm m$^{-2}$ \citep[][]{AlObaidiEtAl2014,TalebAntony2020}, a standard value 
for ground-level direct solar illuminance that is consistent with the top-of-the-atmosphere 
illuminance of 133,600 lm m$^{-2}$ deduced from the 1985 Wehrli Standard Extraterrestrial 
Solar Irradiance Spectrum \citep[][]{NREL2021,Wehrli1985,NeckelLabs1981}. The scattering 
cross-section of a spherical object asymptotically approaches $C_{sca} = \pi r^{2} Q_{sca}(r)$
with the average value for all objects 
\begin{eqnarray}\label{eq:crosssection}
    \sigma &=& 2 \pi \int_{5 \times 10^{-7} \textrm{~m}}^{5 \textrm{~m}} r^{2} R(r) dr~
\end{eqnarray}
and the contribution from the space objects to the zenith luminance is
\begin{eqnarray}\label{eq:luminance}
    L &=& \sigma \alpha \frac{E_{0,vis}}{4\pi} \int_{2 \times 10^{5} \textrm{~m}}^{4 
    \times 10^{7} \textrm{~m}}
	H(h) dh
	\nonumber \\
	&=& \frac{E_{0,vis} \alpha}{2} \int_{5 \times 10^{-7} \textrm{~m}}^{5 \textrm{~m}} 
	r^{2} R(r) dr
	\int_{2 \times 10^{5} \textrm{~m}}^{4 \times 10^{7} \textrm{~m}} H(h) dh
	\nonumber \\
	&\approx& \alpha ~7.2 ~\mu\textrm{cd~m}^{-2},
\end{eqnarray}
where $\alpha$ is the average albedo of space objects. 

Due to their different shapes and orientations, the reflectivity of space objects can vary
substantially. Many spacecraft are made of composite materials, much of them 
highly reflective for the benefit of thermal management, but other objects can appear dark
because of efficient absorption. We use the average value of $\alpha\approx$ 0.5 in 
accordance with \citet[][]{KrutzEtAl2011}. 

\begin{figure}
	\includegraphics[width=\columnwidth]{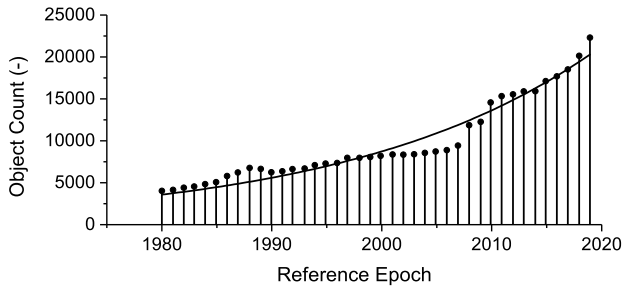}
    \caption{Trend of the absolute number of all counted space objects. Vertical lines 
    with dots represent data provided by \citet[][]{ESA2020}. The solid line is an 
    analytical model.}
   \label{fig:trend}
\end{figure}

Note that the estimate in Eq.~(\ref{eq:luminance}) is specifically applicable to 
conditions during the middle to late 1990s. Based on the trend published by 
\citet[][]{ESA2020} (Fig.~\ref{fig:trend}) we found that the number of known space objects
has increased by a factor of 4.5 since the late 1990s. Therefore, we expect that today's 
contribution from space objects to zenith luminance is at least (7.2$\alpha$) $\times$ 4.5
= 16.2 $\mu$cd m$^{-2}$. Assuming the above increase rate remains conserved in next few 
years, the luminance may quickly approach 25 $\mu$cd m$^{-2}$ in 2030.

\section{Discussion}
\label{sec:results}

Any piece of matter in Earth orbit illuminated by the Sun reflects or
scatters light, and can in principle be detected and tracked as an individual moving
object if it can be distinguished from neighboring ones and its radiance is above  
the sensitivity threshold of one's detector. If its angular size is smaller than the instrument point-spread function 
(PSF), what is recorded in the image is essentially the PSF with an average
irradiance directly proportional to the object radiance and inversely proportional to 
the PSF area. The irradiance detection threshold can be reached either by 
increasing the telescope numerical aperture, yielding larger 
irradiances, and/or by increasing the responsivity 
of the detector. Most present-day instruments used for narrow-field imaging or for large
astronomical surveys have enough performance as to record the individual streaks 
produced by many of the objects orbiting Earth, excepting those of very small sizes.

The situation is different when observing the night sky with the unaided eye. The human 
eye has a PSF with perceptual angular resolution of order 1\arcmin, but it has a 
relatively small light sensitivity in comparison with long exposure images, even under
scotopic adaptation. Except for the brighter ones, most Earth-orbiting objects cannot 
be visually detected and tracked individually, since their individual irradiances fall 
below the visual detection threshold. However, if several such objects are present within the receptive field of a retinal ganglion cell, their combined irradiance may well reach the threshold, and may be perceived as a diffuse skyglow component. 

A similar diffuse skyglow effect arises when measuring the night sky brightness with
highly sensitive but very low angular resolution detectors. Examples of these devices are 
the widely used Sky Quality Meter~(SQM;~\citealt[][]{Cinzano2007}) and Telescope Encoder 
and Sky Sensor-WiFi (TESS-W; \citealt{Zamorano2017}), whose PSFs have full widths at 
half-maximum of $\sim$20$^{\circ}$. In that case the radiance of even very dim space 
objects may be enough to be individually detected, but the low angular resolution of the 
instrument unavoidably integrates the light of the many such objects present within its 
field of view without resolving each individual source. This results in an increased value
of the diffuse night sky brightness, similar to the one that would be produced by light 
pollution from atmospheric scattering of the radiance emitted by anthropogenic light 
sources on the ground.

The estimated strength of this effect, as per direct application of
Eq.~(\ref{eq:luminance}), is currently of order 16.2 $\mu$cd m$^{-2}$. Note that this 
value was obtained from Eq.~(\ref{eq:luminance}) by integrating the contributions of the 
objects located toward the observer's local zenith with a lower integration limit of 200 
km, consistent with the object altitude distribution in Fig.~\ref{fig:stratification} 
and assuming that only one half of the cross-section of the object is illuminated by
sunlight as seen from the observer's position. This corresponds to the objects' phase 
angle at sunset or sunrise. 

At the beginning or the end of the astronomical night, when the Sun is at  
–18$^{\circ}$ with respect to the horizon, the Earth shadow reaches 328 km above the observer. Up to this orbital altitude the space objects are entirely in darkness at these moments. However, since such low-altitude orbits are relatively little 
populated (Fig.~\ref{fig:stratification}) the total number of illuminated space objects is still 99.98 per cent of the maximum. Furthermore, at that time of the night the phase angle of the objects located above the observer is $\pm$72$^{\circ}$, 
so that the fraction of their illuminated surface is 0.65, instead of 0.5. The skyglow 
contribution of the space object cloud at the beginning and end of the astronomical night 
is then proportionally increased, reaching an estimated value of 21.1 $\mu$cd m$^{-2}$.

These luminances amount to $\sim$10 per cent of the natural night sky brightness, a 
critical level mentioned in the 1979 resolution of the International Astronomical Union 
(IAU) as the limiting acceptable value of light pollution at astronomical observatory 
sites. According to this criterion, ``\ldots the increase in sky brightness at 
45$^{\circ}$ elevation due to artificial light scattered from clear sky should not exceed 
10 per cent of the lowest natural level in any part of the spectrum between wavelengths 
300 and 1000 nm except for the spectral line emission from low pressure sodium 
lamps\ldots'' \citep[][]{Cayrel1979}. Although the original criterion refers to the 
brightness at 45$^{\circ}$ elevation, it is reasonable to apply it also to zenith 
observations. 

The concept of a `natural level' of brightness is in itself debatable, since the natural 
night sky brightness in any observation band, including the visual one, is 
widely variable depending on the region of the sky, the location of the observer,
the state of the atmosphere and the highly fluctuating strength of the airglow, as 
described in, e.g., \citet[][]{MasanaEtAl2021} and references therein. According to the 
Gaia-Hipparcos map, zenith luminance values of 200 $\mu$cd m$^{-2}$ could be taken as a 
reasonable `dark sky' luminance benchmark. A reference level of 22.0 mag arcsec$^{-2}$ in 
the Johnson $V$ band is often quoted in the literature, subject to the same variability 
factors as the luminance value. 

Note that the true visual luminance of a Johnson $V$ = 22.0 mag arcsec$^{-2}$ sky 
is contingent on the arbitrarily chosen definition of the $V$ magnitude scale (AB or Vega 
zero-points, Vega magnitude, etc.), and also on the  spectrum of the incoming light. The 
same Johnson $V$ magnitude may correspond to very different sky luminances depending on 
the correlated colour temperature (CCT) of the sky spectrum, as reported in 
\citet[][]{BaraEtAl2020}. Notwithstanding that, 22.0 Johnson $V$ mag arcsec$^{-2}$ corresponds 
to $\sim$200 $\mu$cd m$^{-2}$ for typical sky CCTs recorded in observatories across the 
world \citep[][]{BaraEtAl2020}, so both quantities can be taken as sensible and practical 
reference values for the brightness of the natural sky.

These results imply that diffuse night sky brightness produced by artificial space objects
directly illuminated by the Sun may well have reached nowadays, and perhaps exceeded, what
is considered a sustainability `red line' for ground based astronomical observatory sites.
Note that the data  used in Figs.~(\ref{fig:cumulative}-\ref{fig:trend}) are for 
detected objects only. This means the real number of objects should be even higher
because, in principle, not all objects have been identified. Therefore, the above estimate
is a lower limit. This effect will certainly be aggravated by the planned deployment 
of huge satellite `mega-constellations' that will add a substantial number of reflecting 
objects in large-inclination orbits at altitudes above $\sim$ 500 km. The effect of these 
space objects on the diffuse brightness of the night sky is to be considered in addition 
to the streaks of individual objects that tend to compromise the quality of astronomical 
images. This is in our opinion an issue that should be taken into account by the 
astrophysical community in its efforts to preserve science-grade quality night skies.

The approach in this work is a first approximation to the problem, made with 
some simplifying assumptions. However, the results in Section 2 were derived from robust 
first principles and we believe they capture the basic physics of this effect. Our 
estimates refer only to the human visual band; estimates in other bands of the optical 
region, including the near-infrared, would be also informative but were not addressed in 
this work.

Observational campaigns to evaluate the strength of this effect 
should be planned and carried out. They pose an interesting methodological problem, since 
this new skyglow component corresponds to a diffuse contribution slowly varying across the
night sky, so no sharp borders between an affected and an unaffected area of the sky are 
to be expected. If these borders between adjacent patches of the sky were present, visual 
inspection could have been enough to detect this phenomenon: the luminance
contrast of a 20 $\mu$cd m$^{-2}$ brighter patch over a 200 $\mu$cd m$^{-2}$ background is
0.1, well above the human detection contrast threshold for a 6$^{\circ}$ zone surrounded 
by a 200 $\mu$cd m$^{-2}$ adaptation luminance field, which is 0.07 
\citep[][]{Blackwell1946}. The lack of such sharp borders should prompt the community to 
make absolute measurements and compare them with models of the natural night sky and/or to
monitor changes in night sky brightness on timescales of order of a few years. 
Comprehensive datasets of the present levels of skyglow in dark sites could provide an 
instrumental baseline to assess the additional contribution of the new satellite 
mega-constellations, and also to monitor its evolution in forthcoming years.

\section{Prospects and Conclusions}
\label{sec:conclusions} 
The cloud of artificial objects orbiting the Earth, composed of both operational and 
decommissioned satellites, parts of launch vehicles, fragments, and small particles, with 
characteristic sizes ranging from micrometers to tens of meters, reflect and scatter 
sunlight toward ground-based observers. When imaged with high angular resolution and high 
sensitivity detectors, many of these objects appear as individual streaks in science 
images. However, when observed with relatively low-sensitivity detectors like the unaided 
human eye, or with low-angular-resolution photometers, their combined effect is that of a 
diffuse night sky brightness component, much like the unresolved integrated starlight 
background of the Milky Way. 

According to our preliminary estimations, this newly recognised skyglow component could 
have reached already a zenith visual luminance of about 20 $\mu$cd m$^{-2}$, which 
corresponds to 10 per cent of the luminance of a typical natural night sky, exceeding in 
that way the IAU's limiting light pollution `red line' for astronomical observatory sites.
Future satellite mega-constellations are expected to increase significantly this light 
pollution source.

\section*{Acknowledgements}

This work was supported by the Slovak Research and Development Agency under 
contract No: APVV-18-0014. Computational work was supported by the Slovak National 
Grant Agency VEGA (grant No. 2/0010/20). SB acknowledges support by Xunta de 
Galicia (grant ED431B 2020/29).

\section*{Data Availability Statement}

Digitized data shown in Fig.~\ref{fig:stratification}-\ref{fig:trend} and the numerical solvers used to model zenith NSB are publicly available on \url{http://davinci.fmph.uniba.sk/~kundracik1/debris}. We did not use any new data.








%
%


\bsp	
\label{lastpage}
\end{document}